\documentclass{IEEEtran}
\usepackage[latin9]{inputenc}
\usepackage{amsmath}
\usepackage{amssymb}
\usepackage{graphicx}
\usepackage{hyperref}

\makeatletter

\providecommand{\tabularnewline}{\\}

\usepackage{cite}
\usepackage{xcolor}

\usepackage{amsfonts}\usepackage{algorithmic}
\def\BibTeX{{\rm B\kern-.05em{\sc i\kern-.025em b}\kern-.08em
    T\kern-.1667em\lower.7ex\hbox{E}\kern-.125emX}}
\PassOptionsToPackage{position=top}{subfig}
\PassOptionsToPackage{justification=raggedright,singlelinecheck=false}{subfig}
\PassOptionsToPackage{caption=false}{subfig}
\usepackage{subfig}
\makeatother

\begin{document}

\title{Metagrating-Assisted High-Directivity Sparse Antenna Arrays For Scanning Applications}

\author{Yaniv Kerzhner and Ariel Epstein, \IEEEmembership{Senior Member, IEEE}
\thanks{Manuscript received XXX, 2022; revised YYY, 2021; accepted ZZZ, 2022.
Date of publication WWW, 2022; date of current version UUU, 2021.
This work was supported by the Israel Science Foundation under Grant
1540/18.} \thanks{Y. Kerzhner and A. Epstein are with are with the Andrew and Erna Viterbi
Faculty of Electrical and Computer Engineering, Technion - Israel Institute of
Technology, Haifa 3200003, Israel (email: yanivk2@campus.technion.ac.il;
epsteina@ee.technion.ac.il).}}
\maketitle
\begin{abstract}
We present an analytical scheme for designing metagrating-enhanced sparse antenna arrays. Unlike previous work, the proposed method does not involve time-consuming cost function optimizations, complex structural manipulations on the active array or demanding computational capabilities. Instead, it merely requires the integration of a passive metagrating (MG) superstrate, a planar periodic arrangement of subwavelength capacitively-loaded wires (meta-atoms), synthesized conveniently via a semianalytical procedure to guarantee suppression of grating lobes in the sparse configuration. Correspondingly, we extend previous formulations to enable excitation of the MG by the active array elements, deriving analytical relations connecting the passive and active element distribution and electrical properties with the scattered fields, eventually allowing resolution of the detailed device configuration leading to optimal directivity. Importantly, considering typical active array applications, the semianalytical synthesis scheme is further developed to take full advantage of the various degrees of freedom in the system, harnessing them to support scanning in a wide range of extreme angles while maintaining a single directive beam. The resultant methodology, verified in simulations to work well also for large finite arrays, offers an original path for mitigating grating lobes in sparse arrays with scanning capabilities, yielding a complete printed-circuit-board compatible design without relying on full-wave optimization.
\end{abstract}


\section{Introduction\label{sec:Introduction}}

Antenna arrays, consisting of multiple (typically equally spaced) radiating elements with individual amplitude and phase control, are well known as effective configurations for achieving desired radiation pattern properties, such as directivity, main beam angle, beam width, and side-lobe level\cite{balanis2016antenna}. In general, interelement spacing must not exceed half the wavelength; otherwise, grating lobes might emerge, deteriorating directivity. On the other hand, setting a small distance between the elements to comply with this constraint leads to increased mutual coupling effects\cite{edelberg1960mutual,qamar2016mutual,kim2019compact,zhang2019mutual}, preventing the use of simplistic antenna array theory. In addition, the larger amount of elements (per unit area) requires a large amount
of power amplifiers, which significantly increases the complexity of the design and may lead to reduced efficiency \cite{cheng2014w,huang2015antenna,mandal2021low}.

In view of these challenges, numerous studies have been conducted along the years, aiming at reducing the element density while retaining the appealing antenna array radiation properties \cite{leahy1991design,zhang2011reducing,bencivenni2014design,Chakravorty2016,Iqbal2019,Xu2019,pinchera2019efficient,Ren2022}. \textcolor{black}{While some authors propose to engineer the element pattern such that radiation towards the array-factor-associated grating lobe would be minimized overall \cite{Chakravorty2016,Iqbal2019,Ren2022}, this approach requires application-specific design of the active radiators, and is not always easy to apply for scanned array scenarios.} \textcolor{black}{Instead,} most of the \textcolor{black}{suggested} methods use unequally spaced and non uniform excitation strategies to overcome the classical element-density/grating-lobes trade-off, and are generally based on optimization of a nonlinear cost function over some constraints which favor sparse arrays \cite{leahy1991design,zhang2011reducing,bencivenni2014design,Xu2019,pinchera2019efficient}. Usually, the solution relies on global optimization methods, giving
rise to \textcolor{black}{three} main issues. First, these kind of methods are typically time consuming and require substantial computational resources. Second, the absence of an analytical model to guide the design process makes it difficult to draw general conclusions regarding the limitations of these methods and the physical mechanism underlying the solution. \textcolor{black}{Lastly, to perform well, designated circuitry to accommodate the irregular active element distribution and specialized excitation profiles is required, which may complicate the overall implementation with respect to the well-established, conventional, uniform array configurations.} 

In this paper, we introduce an alternative approach to design sparse antenna arrays, based on the concept of metagratings (MGs) - sparse periodic arrangements of subwavelength polarizable particles (meta-atoms),
engineered to exhibit desired scattering properties \cite{ra2021metagratings}. Unlike metasurfaces, in
which the meta-atoms are densely packed to allow abstract (homogenized) design based on the generalized sheet transition conditions \cite{kuester2003averaged,tretyakov2003analytical},
MGs are not synthesized following the homogenization approximation.
As a consequence, the meta-atoms can be sparsely distributed in space, and the design relies on analytical models, reliably accounting for near- and far- field coupling between the particles\cite{ra2017metagratings}.
Since these models consider the actual MG layout (and not an abstract constituent distribution, as in metasurface design \cite{epstein2016huygens}), many times even the particle geometries, and since MGs typically contain
far less meta-atoms per period, the detailed design of a practical working prototype becomes considerably simpler. 

Indeed, in recent years, extensive research on MGs revealed their efficiency in implementing beam-manipulation functionalities \cite{ra2017metagratings,memarian2017wide,sell2017large,wong2018perfect,epstein2017unveiling,epstein2018perfect,yang2018freeform,rabinovich2018analytical,popov2018controlling}. In particular, at microwave frequencies, semianalytical \textcolor{black}{methodologies} directly yielding fabrication-ready design specifications have been devised, subsequently used to demonstrate (theoretically and experimentally) printed-circuit-board (PCB) MGs for perfect anomalous reflection \cite{rabinovich2018analytical,rabinovich2019experimental,Xu2021dual},
refraction \cite{epstein2018perfect}, waveguide mode manipulations \cite{killamsetty2021metagratings,biniashvili2022perfect}, and arbitrary diffraction engineering \cite{casolaro2019dynamic,rabinovich2020arbitrary,Popov2020Beamforming,popov2021non,xu2021analysis}.
Since MGs clearly excel in suppression spurious Floquet-Bloch (FB) modes, and funneling all the incoming power to specified directions in space, it would only be natural to try and harness this innovative concept to develop an efficient spatial filter for eliminating grating lobes in sparse antenna arrays \cite{kerzhner2019suppressing}. Besides the appealing synthesis scheme, such a solution may enable leaving
the original active antenna array design unchanged, while improving substantially its performance with the modular addition of the passive MG.

\medskip{}

\begin{figure*}[!t]
\subfloat[]{\includegraphics[width=0.6\textwidth]{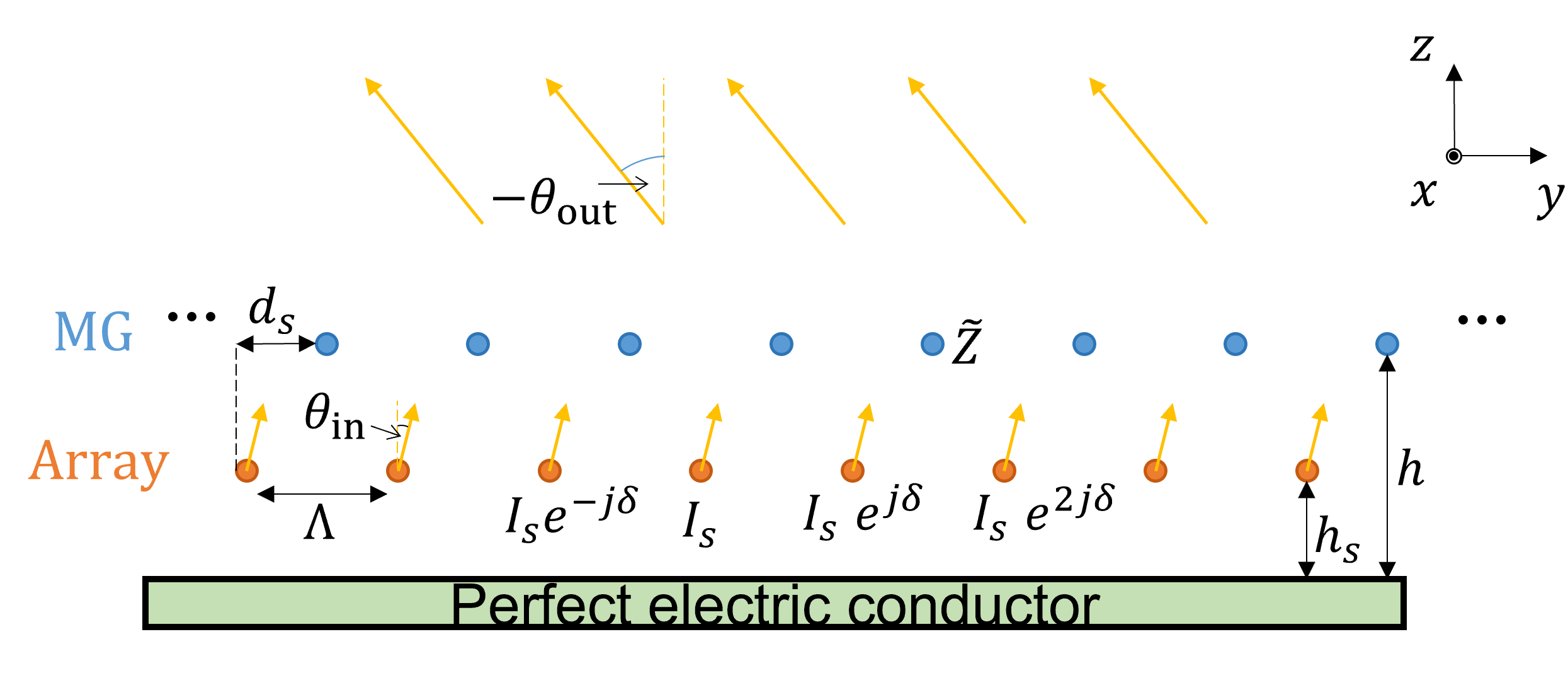}

}\hfill{}\subfloat[\label{fig:single-cap}]{\includegraphics[width=0.35\textwidth]{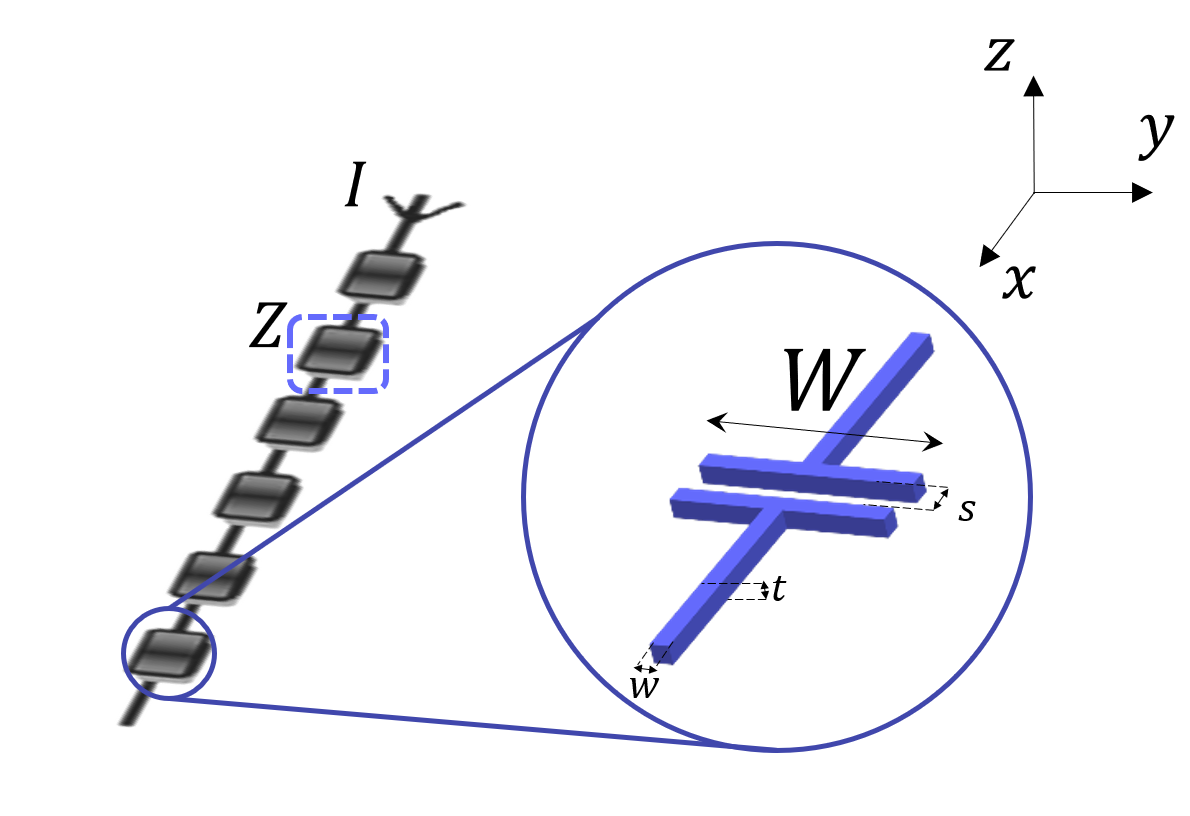}

}

\caption{Physical configuration of the sparse antenna array, loaded by a MG
superstrate backed by a perfect electric conductor. (a) Front view
of the $\Lambda$periodic system, composed of active electric line
sources array with a $\delta$ phase delay between its elements, and
a passive capacitively-loaded conducting wires separated by $h_{s}$
and $h$ respectively from a PEC with a $y$-direction offset of $d_{s}$
between them. All the energy is coupled to the desired angle $\theta_{\text{out}}$,
while eliminating all the side lobes (b) A single passive capacitively-loaded
conducting wire with a distributed impedance per unit length of $\widetilde{Z}$,
formed by periodic structure of capacitors along the x axis. Each
capacitor has trace width of $w$, and the load impedance is controlled
by the capacitor width $W$. \label{fig:system}}
\end{figure*}

Correspondingly, we propose herein a radiating device composed of a sparse antenna array backed by a ground plane, covered with a passive MG superstrate featuring capacitively loaded wires as meta-atoms (Fig. \ref{fig:system}). As so far MGs have mainly been used in configurations where plane-wave or beam excitation was considered, we extend the analytical model previously developed by us \cite{epstein2017unveiling} to allow integration of the localized active array elements into the
synthesis scheme. Once such a theoretical model is established, we utilize it to demonstrate sparse antenna arrays with fixed beam and scanning capabilities, featuring negligible coupling to spurious lobes, validating the results in a full-wave solver. Similar to previous MGs works \cite{epstein2017unveiling,rabinovich2018analytical}, utilizing the model we retrieve the positions of the array and the MG relative to each other and the geometry of the MG capacitors that would eliminate spurious FB modes while conserving the power, leading to a passive and lossless design. Subsequently, acknowledging based on the analytical model that additional degrees of freedom are available for utilization, we further extend the methodology to enable MG designs accommodating multiple angles of incidence simultaneously, thus facilitating grating-lobe-free scanning functionalities. Finally, we examine the performance of the resultant design in a more realistic scenario, where a finite array is considered, and show that the scheme remains highly effective even in this case. The results yield an efficient synthesis procedure for enhanced antenna arrays, enabling not only dramatic sparsity, but also an expansion of the scanning range. Importantly, they further demonstrate the versatility and usefulness of the metagrating concept, promoting its integration into antenna devices.

\section{Theory\label{sec:Theory}}

\subsection{Formulation}

We consider a 2D $\left(\partial/\partial x=0\right)$ $\Lambda$-periodic
configuration composed of $active$ electric line sources (the antenna
array), covered with $passive$ capacitively-loaded conducting wires
(the MG superstrate), backed by a perfect electric conductor (PEC)
layer at $z=0$ (Fig. \ref{fig:system}). The passive conducting strips
are of width $w\ll\lambda$ and thickness $t\ll\lambda$, where $\lambda$
is the wavelength. Due to the subwavelength periodicity $L\ll\lambda$
along the x direction, these strips are assumed to be uniformly loaded
by impedance per unit length $\tilde{Z}$\cite{tretyakov2003analytical,tretyakovPavel2003loadedWire,ikonen2007modeling,epstein2017unveiling}.
The distances from the active array and the passive MG to the PEC
are denoted by $h_{s}$ and $h$ respectively, and the horizontal
offset between them is $d_{s}$. 

The system is excited by the current in the active lines following
the typical uniform array scheme\cite{balanis2016antenna}, with a
magnitude of $I_{s}$ and a phase shift of $\delta=-2\pi\left(\Lambda/\lambda\right)\sin\left(\theta_{\text{in}}\right)$
between the elements such that the entire active array would, in principle,
radiate at an angle of $\theta_{\text{in}}$ with respect to the $z$
axis. We are interested in cases where the array is sparse, such that
there exists a single grating lobe directed to an angle denoted as
$\theta_{\text{out}}$, which would usually cause performance impairment
(reduced directivity). Our objective is to design the overall system such that all the power will be radiated
towards a single angle $\theta_{\text{out}}$, without any
other main or grating lobes, despite its sparsity. To this end, we
utilize the configuration\textquoteright s four degrees of freedom,
namely, $h,h_{s},d_{s}$ and $\tilde{Z}$, to guarantee exclusive coupling to a single radiating mode.

As in \cite{epstein2017unveiling}, we start by expressing the total
field in space as a superposition of the field generated by the impressed
currents $I_{s}e^{jn\delta}$ in the active array, the induced $Ie^{jn\delta}$
in the passive MG, and their images due to the PEC mirror,  
\begin{equation}
\begin{array}{l}
E_{x}^{\text{total}}\left(y,z\right)=\\
\,\,-\frac{k\eta}{4}\!\!\!\sum\limits _{n=-\infty}^{\infty}\!\!\!\!I_{s}e^{jn\delta}\!\!\left\{\!\!\!\! \begin{array}{l}
H_{0}^{\left(2\right)}\!\!\left[k\sqrt{\left(y-d_{s}-n\Lambda\right)^{2}+\left(z-h_{s}\right)^{2}}\right]\\
\,-H_{0}^{\left(2\right)}\!\!\left[k\sqrt{\left(y-d_{s}-n\Lambda\right)^{2}+\left(z+h_{s}\right)^{2}}\right]
\end{array}\!\!\!\!\right\} \vspace{3 pt} \\ 
\,\,-\frac{k\eta}{4}\!\!\!\sum\limits _{n=-\infty}^{\infty}\!\!\!\!Ie^{jn\delta}\!\!\left\{ \!\!\!\!\begin{array}{l}
H_{0}^{\left(2\right)}\!\!\left[k\sqrt{\left(y-n\Lambda\right)^{2}+\left(z-h\right)^{2}}\right]\\
\,-H_{0}^{\left(2\right)}\!\!\left[k\sqrt{\left(y-n\Lambda\right)^{2}+\left(z+h\right)^{2}}\right]
\end{array}\!\!\!\!\right\} 
\end{array} \label{eq:total-field-Hankel}
\end{equation}
where $H_{0}^{\left(2\right)}\!\left(\Omega\right)$ is the zeroth-order Hankel function of the second kind, $k=2\pi/\lambda$ is the
wavenumber, and $\eta=\sqrt{\mu/\epsilon}$ is the wave impedance.
Since the configuration is effectively invariant along $x$, and the
excitation currents flow parallel to the $x$ axis as well, the resultant
fields are transverse electric (TE) polarized, with $E_{y}=E_{z}=H_{x}=0$.

Using the Poisson formula and the Fourier transform of the Hankel function we express the fields as a modal FB sum \cite{marcuvitz1973radiation,tretyakov2003analytical},
\begin{equation}
\begin{array}{l}
E_{x}^{\text{total}}\left(y,z\right)= 
-\frac{k\eta}{2\Lambda}\!\!\!\!\sum\limits _{m=-\infty}^{\infty}\!\!\!\!E_{m}\textcolor{black}{(y,z)}
\end{array}
\label{eq:total-field}
\end{equation}
where $E_{m}\textcolor{black}{(y,z)}$ is given by
\begin{equation}
\begin{array}{l}
\!\!\!\!\!\!\!E_{m}\left(y,z\right)= \vspace{3 pt} \\ 
\frac{e^{-jk_{t_{m}}y}}{\beta_{m}}\!\! \left[\!\!\!\begin{array}{l}
I\left(e^{-j\beta_{m}\left|z-h\right|}-e^{-j\beta_{m}\left|z+h\right|}\right) \\
+e^{jk_{t_{m}}d_{s}}I_{s}\left(e^{-j\beta_{m}\left|z-h_{s}\right|}-e^{-j\beta_{m}\left|z+h_{s}\right|}\right)
\end{array}\!\!\!\right] \vspace{3 pt}
\end{array}
\label{eq:Em}
\end{equation}
and where the transverse and longitudinal wavenumbers, respectively, are $k_{t_{m}}=\frac{2\pi m}{\Lambda}+k\sin\theta_{\text{in}}$ and $\beta_{m}=\sqrt{k^{2}-k_{t_{m}}^{2}}$ ($\Im\left\{ \beta_{m}\right\}<0$). As can be deduced from (\ref{eq:total-field}), since the theory is developed for the case of infinite periodic arrays, the fields can be described as a discrete set of FB modes, with only a finite number of them propagating (corresponding to far-field radiation from the system). However, it is clear that eventually a finite array will be employed in practice; the implications of this truncation will be discussed in detail in Subsection \ref{subsec:Finite-Arrays}.

\subsection{Propagating mode selection rules}
To be able to address the problem with a simple MG, featuring a single
element per period, we wish to consider the simplest case possible
that still allows sparsity beyond the conventional harsh $\Lambda<\lambda/2$
separation restriction. In particular, to guarantee that only one
grating lobe exists, the interelement spacing must satisfy\cite{rabinovich2018analytical},
\begin{equation}
\frac{\lambda}{1+\sin\theta_{\text{in}}}<\Lambda<\min\left\{\frac{\lambda}{1-\sin\theta_{\text{in}}}, \frac{2\lambda}{1+\sin\theta_{\text{in}}}\right\}
\label{eq:Lambda_restriction}
\end{equation}
for given phasing $\delta$ (given \textquotedblright angle of incidence\textquotedblright{}
$\theta_{\text{in}}\in\left(0,\pi/2\right)$). Since we intend to use the MG superstrate
also as a scan-range extender, we wish to select an appropriate $\theta_{\text{out}}$ for the eventually radiated main beam. The rationale is to enable small phase shifts $\delta$ in the active array (corresponding to small $\theta_{\text{in}}$) to be translated to radiation towards a grazing angle. Hence, the interelement spacing should be chosen such that
\begin{equation}
\Lambda=\frac{\lambda}{\left|\sin\theta_{\text{in}}-\sin\theta_{\text{out}}\right|}\label{eq:Lambda}
\end{equation}
under the restrictions of (\ref{eq:Lambda_restriction}). In this case, in the observation region $z>h$, the $m=0$ mode in (\ref{eq:total-field}) would correspond to radiation towards $\theta_{\text{in}}$, the $m=-1$ mode will radiate towards $\theta_{\text{out}}$, and the rest of the modes will be evanescent (will not contribute to the far fields).

\subsection{Spurious lobe suppression}

With the chosen phasing $\delta$ and interelement spacing $\Lambda$,
the array current excitation could lead to radiation towards either
$\theta_{\text{in}}$ or $\theta_{\text{out}}$ (or both). To maximize
directivity while steering the beam to the large oblique angle $\theta_{\text{out}}$,
we should first require that no power will be coupled to the spurious
lobe at $\theta_{\text{in}}$. This is achieved by requiring the corresponding (m=0) scattering coefficient in \eqref{eq:total-field} to vanish in the observation region $z>h$.
This constraint implies that in order to suppress the undesired grating
lobe, we should set the configuration's degrees of freedom such that
the induced current in the passive array will follow
\begin{equation}
I=-I_{s}e^{jkd_{s}\sin\theta_{\text{in}}}\frac{\sin\left(kh_{s}\cos\theta_{\text{in}}\right)}{\sin\left(kh\cos\theta_{\text{in}}\right)}
\label{eq:current}
\end{equation}

As in \cite{epstein2017unveiling,rabinovich2018analytical,rabinovich2020arbitrary},
to ensure the proper current will indeed be induced on the MG upon
excitation by the active array, we need to suitably tune the load
impedance as to satisfy Ohm's law $E_{x}^{\text{tot}}\left(y\rightarrow0,z\rightarrow h\right)=\tilde{Z}I$
on the reference wire\cite{tretyakov2003analytical}. Based on the
scattering analysis for a narrow thin metal strip, we deduce that
the induced current in the flat wire of width $w$ is equivalent to the
one that would appear on a conducting cylinder with a radius of $r_{\text{eff}}=w/4$
\cite{barkeshli2004electromagnetic}. Utilizing the Hankel function asymptotic approximation for small arguments\cite{abramowitz1965handbook} and Bessel function sum identities\cite{gradshteyn2015table} in \eqref{eq:total-field-Hankel}-\eqref{eq:total-field} and substituting \eqref{eq:current}, the distributed load impedance leading to optimal directivity reads \cite{epstein2017unveiling,rabinovich2018analytical,rabinovich2020arbitrary,tretyakov2003analytical} 
\begin{equation}
\begin{array}{l}
\tilde{Z}= -\frac{j\eta}{\lambda}\log\left(\frac{2\Lambda}{\pi w}\right)-k\eta\frac{1-e^{-2j\beta_{0}h}}{2\Lambda\beta_{0}} \\
\,\, -k\eta\sum\limits _x\left(\frac{1-e^{-2j\beta_{m}h}}{2\Lambda\beta_{m}}-\frac{j}{4\pi\left|m\right|}\right)\\
\,\,+k\eta\frac{\sin\left(\beta_0 h\right)}{\sin\left(\beta_0 h_{s}\right)}\sum\limits _{m=-\infty}^{\infty}\frac{e^{-j\beta_{m}\left|h-h_{s}\right|}-e^{-j\beta_{m}\left(h+h_{s}\right)}}{2\Lambda\beta_{m}}e^{j\frac{2\pi m}{\Lambda}d_{s}}
\end{array}
\label{eq:impedance}
\end{equation}

\subsection{Passive lossless superstrate}

Clearly, we wish to realize the MG using the passive configuration
of Fig. \ref{fig:system}, and in particular using printed capacitors
as loads\footnote{Inductive loads can be readily realized as well using standard PCB technology, as printed meander lines \cite{popov2019designing}. However, we prefer herein to restrict ourselves to capacitive loads, which were found to be sufficient for all considered case studies, since they allow convenient utilization of our previously derived semianalytical approximation tying the printed capacitor dimensions and the effective load impedance \cite{epstein2017unveiling}, presented and used in Section \ref{subsec:Fixed-beam-phased-arrays}.}.
Therefore, an additional constraint on the design parameters is
required, ensuring that the evaluated load impedance per unit length
\eqref{eq:impedance} is indeed reactive, reading
\begin{equation}
\Re\left\{ \tilde{Z}\right\} =0\label{eq:real-part-Z}
\end{equation}
This is different from the passivity conditions previously formulated for beam manipulating MGs, e.g. for anomalous reflection  \cite{epstein2017unveiling, rabinovich2018analytical}, where the total power coupled to the radiated mode ($m=-1$ in the considered case) was coerced to coincide with the incident power. The reason for this difference lies in the different nature of the excitations under consideration. When a plane wave excites the MG, the incident field amplitude allows unambiguous estimation of the impinging power interacting with the structure; however, when the excitation is composed of impressed current sources as considered herein, the source-generated power cannot be evaluated without prior knowledge of its near-field environment (close-by scatterer arrangement, for instance) \textcolor{black}{\cite{krasnok2015antenna}}, which is yet to be determined. Hence, the alternative condition \eqref{eq:real-part-Z} is formulated to ensure MG passivity, similar to \cite{epstein2018perfect}, which does not involve the input power directly.
%

Substituting   \eqref{eq:impedance} into   \eqref{eq:real-part-Z}
we derive a more explicit version of this constraint, 
\begin{equation}
\begin{array}{l}
\!\!\!\!\frac{\sin\left(\beta_{-1}h_{s}\right)}{\beta_{-1}}\left[
\sin\left(\beta_{-1}h\right)\frac{\sin\left(\beta_0 h_{s}\right)}{\sin\left(\beta h\right)}-\sin\left(\frac{2\pi}{\Lambda}d_{s}+\beta_{-1}h\right)
\right]\\
\,\,+\!\!\!\!\sum\limits_{{\scriptsize  \begin{array}{c} {m\!=\!-\infty}\\  {m\!\neq\!0,-1}\end{array}}}^{\infty}\!\!\!\!e^{\alpha_{m}h}\frac{\sinh\left(\alpha_{m}h_{s}\right)}{\alpha_{m}}\sin\left(\frac{2\pi m}{\Lambda}d_{s}\right)=0
\end{array}
\label{eq:condition}
\end{equation}
formulating the required relations between the horizontal and vertical offsets ($h$, $h_{s}$, and $d_{s}$) to obtain the desired functionality \textcolor{black}{[}we use $\beta_{m}\triangleq-j\alpha_{m}$ $\left(\alpha_{m}\geq0,\,\forall m\neq 0, -1\right)$ to define the decay constants in the terms corresponding to evanescent modes\textcolor{black}{]}.
In other words, every combination ($h,h_{s},d_{s}$) that satisfies \eqref{eq:condition} defines a valid array+MG configuration in which a passive and lossless MG could optimally restore the directivity of the antenna array (despite being sparse), provided that the load
impedance follows \eqref{fig:Normalized Ploss} with this chosen set of array and MG element coordinates. Lastly, in order to realize these capacitive loads in practice, we need to convert their electrical characteristics to physical properties that can be used to define the trace layout for manufacturing. Following the methodology developed in \cite{epstein2017unveiling,epstein2018perfect},
the resolved distributed impedance $\tilde{Z}$ can be emulated by
printed capacitors repeating with a subwavelength period along $x$,
with a suitable capacitor width $W$ realizing the required capacitance
per unit length (Fig. \ref{fig:single-cap}); this will be addressed promptly in Section \ref{subsec:Fixed-beam-phased-arrays}.

\section{Results And Discussion}
\label{sec:results}
\subsection{Fixed-beam phased arrays \label{subsec:Fixed-beam-phased-arrays}}

To demonstrate and verify the above synthesis scheme, we follow the
prescribed methodology to design a sparse antenna array, with a $\Lambda=0.93\lambda>0.5\lambda$
interelement spacing (periodicity) along $y$ at an operating frequency of $f=20\text{\,GHz}$. The array will be designed such that for a phase delay of $\delta=-0.98\,\text{rad}$ between its active elements, corresponding to $\theta_{\text{in}}=10^{o}$, all the radiated power will be funnelled into out $\theta_{\text{out }}=-63.93^\circ$, without any spurious grating lobes.

\begin{figure}
\centerline{\includegraphics[width=7 cm]{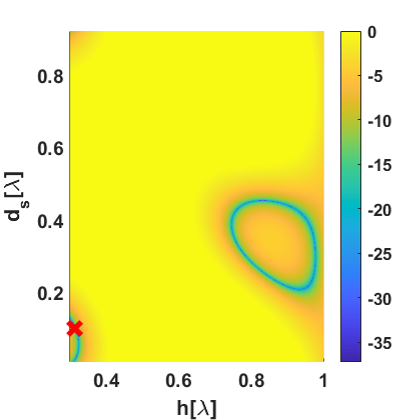}}
\caption{An example of a resulted normalized power loss in $\frac{P_{Loss}}{P_{tot}}\left[dB\right]$
due to the real part of the required $\tilde{Z}$ \eqref{eq:real-part-Z}
to eliminate the 0'th mode \eqref{eq:current}, as function of the vertical distance $h$ of the passive array from the PEC and an offset value $d_{s}$ between the two arrays for an arbitrary active array height value of $h_{s}=0.3\lambda$. The $h$ values consider are only $h>h_{s}$ to fit in to our model. The lossless branches (which corresponding to $\Re\left\{ \tilde{Z}\right\} =0)$ can be seen clearly
as the blue (low values) branches, and an arbitrary coordinate that
we work with is marked by a red x. \label{fig:Normalized Ploss}}
\end{figure}

As can be deduced from Section \ref{sec:Theory}, for the required
functionality the synthesis formalism yields a single nonlinear equation \eqref{eq:condition} tying the three degrees of freedom available in our system ($h$, $h_{s}$ and $d_{s}$). Therefore, the configuration includes two redundant degrees of freedom, which can be chosen at will, in principle. For the purpose of our demonstration, we arbitrarily set the distance between the active array and the PEC to be $h_{s}=0.3\lambda$, and utilize \eqref{eq:impedance} to calculate the deviation from the passivity condition \eqref{eq:real-part-Z}-\eqref{eq:condition} as a function of the remaining degrees of freedom, i.e. the vertical \textcolor{black}{and horizontal} offsets $h$ and $d_{s}$. Figure \ref{fig:Normalized Ploss} presents this deviation using a more physically meaningful figure of merit, as the power required to be absorbed (loss) or provided (gain) to the system by the MG $P_{\mathrm{Gain/Loss}}=\frac{1}{2}\left|\Re\{\tilde{Z}\}\right|\left|I\right|^{2}$ associated with the real part of the calculated $\tilde{Z}\left(h,h_{s},d_{s}\right)$, normalized to the total power input by the active array source $P_{\mathrm{tot}}=P_{\mathrm{rad}}+P_{\mathrm{Gain/Loss}}$, where $P_{\mathrm{rad}}=\sum\limits _{\textcolor{black}{m=-1}}^{\textcolor{black}{0}}\textcolor{black}{\Lambda} \frac{\left|E_{m}\textcolor{black}{(y,z_p)}\right|^{2}}{2\eta_{m}} $,
\textcolor{black}{with} $E_{m}\textcolor{black}{(y,z)}$ \textcolor{black}{of \eqref{eq:Em} evaluated on some plane $z_p>h$} and \textcolor{black}{the modal wave impedance of the $m$th mode defined as} $\eta_{m}=\frac{k\eta}{\beta_{m}}$.

In the graph, plotted in decibel units, vertical and horizontal offsets that enable
implementation of the desired functionality with a passive lossless
MG correspond to dark blue regions, denoting solution branches of
\eqref{eq:condition}. 
Any point on these branches may serve as a suitable coordinate set for placing our meta-atoms. For our demonstration, we choose the point $\left(h,d_{s}\right)=\left(0.314\lambda,0.102\lambda\right)$
marked in Fig. \ref{fig:Normalized Ploss} by a red
cross. Substituting these geometrical parameters into \eqref{eq:impedance}
yields the required load impedance, $\tilde{Z}=-5.53j\left[\eta/\lambda\right]$, which is indeed purely capacitive. 

Finally, in order to obtain the detailed trace geometry featuring
the meta-atom\textquoteright s physical layout, we utilize our previously devised capacitive load configuration \cite{epstein2017unveiling}, with the periodicity along the $x$ axis set to $L=0.1\lambda\ll\lambda$, the trace width and separation to $w=s=3\text{\,mil}$, and the metal thickness to $t=18\,\text{mil}$ \textcolor{black}{(see inset of Fig. \ref{fig:single-cap})}.
To finalize the detailed design, the capacitor width $W$ should be determined as well; following our previous work \cite{epstein2017unveiling}, the required $W$ to realize the distributed impedance $\tilde{Z}$ prescribed by the formulation can be approximated by
\begin{equation}
W=2.85K_{\text{corr}}C\,[\text{mil/fF}]
\label{eq:capWidth}
\end{equation}
where the capacitance is given by $C=-1/\left(2\pi fL\Im\left\{ \tilde{Z}\right\} \right)$ and the correction factor $K_{\text{corr}}$ was evaluated at $f=20\text{\,GHz}$, for the above physical values of $w,s,t$ as $K_{\text{corr}}=0.89$ \cite{epstein2017unveiling}. Utilizing these relations we extract the required printed capacitor width, corresponding in this case to $W=96.93\,\text{mil}$.

\begin{figure}
\centerline{\includegraphics[width = 7 cm]{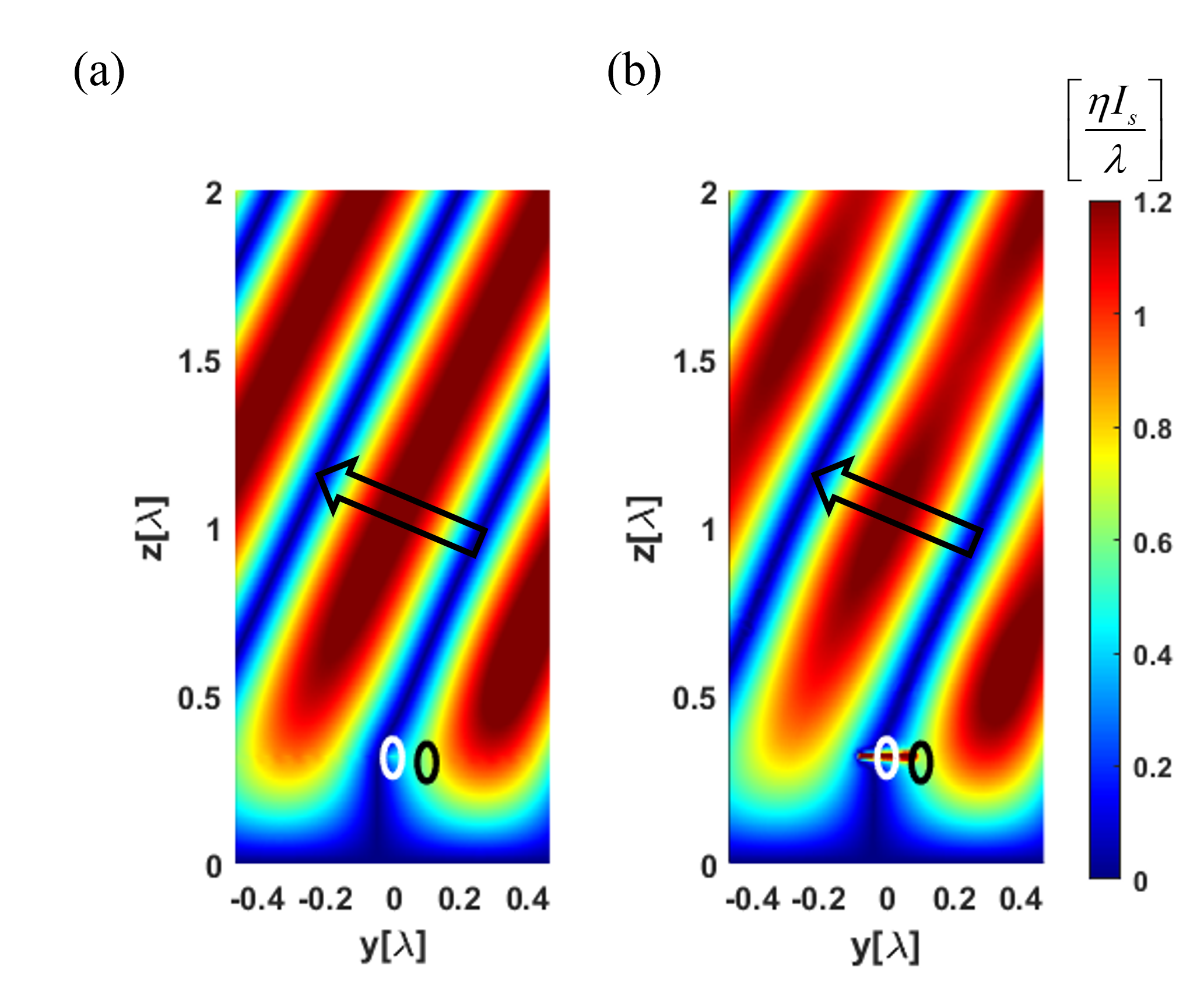}}

\caption{Electric-field distribution $\left|\Re\left\{ E_{x}\left(y,x\right)\right\} \right|$
over a single $\Lambda$-sized period as obtained from the (a) analytical
model and (b) the ANSYS HFSS simulation of the designed sparse antenna
array, fully coupling the original $\theta_{\text{in}}=10^{\circ}$
radiating active array's energy to the desired $\theta_{\text{out}}=-63.93^{\circ}$
direction. \label{fig:Electric-field}}
\end{figure}

The resultant sparse MG-covered array configuration was subsequently defined in a commercial finite element solver, ANSYS HFSS, and simulated under periodic boundary conditions. Realistic copper conductivity of $\sigma=5.8\times10^{7}\left[\text{S/m}\right]$ was used for the metallic traces forming the MG, and current source excitations form the active array. Figure \ref{fig:Electric-field} compares
the fields excited by the active array in the presence of the MG superstate
as received from full-wave simulation to the ones predicted by the
analytical model using \eqref{eq:total-field}. As seen, the field
snapshots agree very well, serving as evidence for the analytical model reliability. Certain discrepancy between the plots can be detected in close proximity to the loaded wire (white circle); however, this is expected due to the finite size of the printed capacitors, producing near-field features that are not reproduced when a distributed load
model is used, as in our analysis \cite{epstein2017unveiling,rabinovich2018analytical}.

Importantly, the field plots clearly indicate that the source excitation is exclusively coupled to a single FB mode, forming highly directive radiation characterized by parallel planar phase fronts. Quantitatively, simulated results imply that more than 99.9\% of the power generated by the active array is coupled towards $\theta_{\text{out}}=-63.93^{\circ}$, with negligible losses dissipated in the MG; as expected, coupling to the spurious grating lobe is indeed negligible, amounting to $\sim0.05\%$ of the main lobe power. In contrast, the sparse array on its own, in the absence of the MG, would radiate as little as $57\%$ of the
power towards $\theta_{\text{out}}=-63.93^{\circ}$, while the rest of the power is radiated towards the acute-angle lobe \textcolor{black}{(Fig. \ref{fig:error_field})}. Clearly, in such a configuration, the array sparsity results in significant gain reduction.
As verified by the \textcolor{black}{HFSS} simulation, the proposed simple passive solution, synthesized based on an analytical model up to the detailed trace geometry, is capable of fully mitigating this effect, without any full-wave optimization whatsoever.

\subsection{Dynamic scanning\label{subsec:Dynamic-scanning}}

\begin{figure}
\centerline{\includegraphics[width = 7 cm]{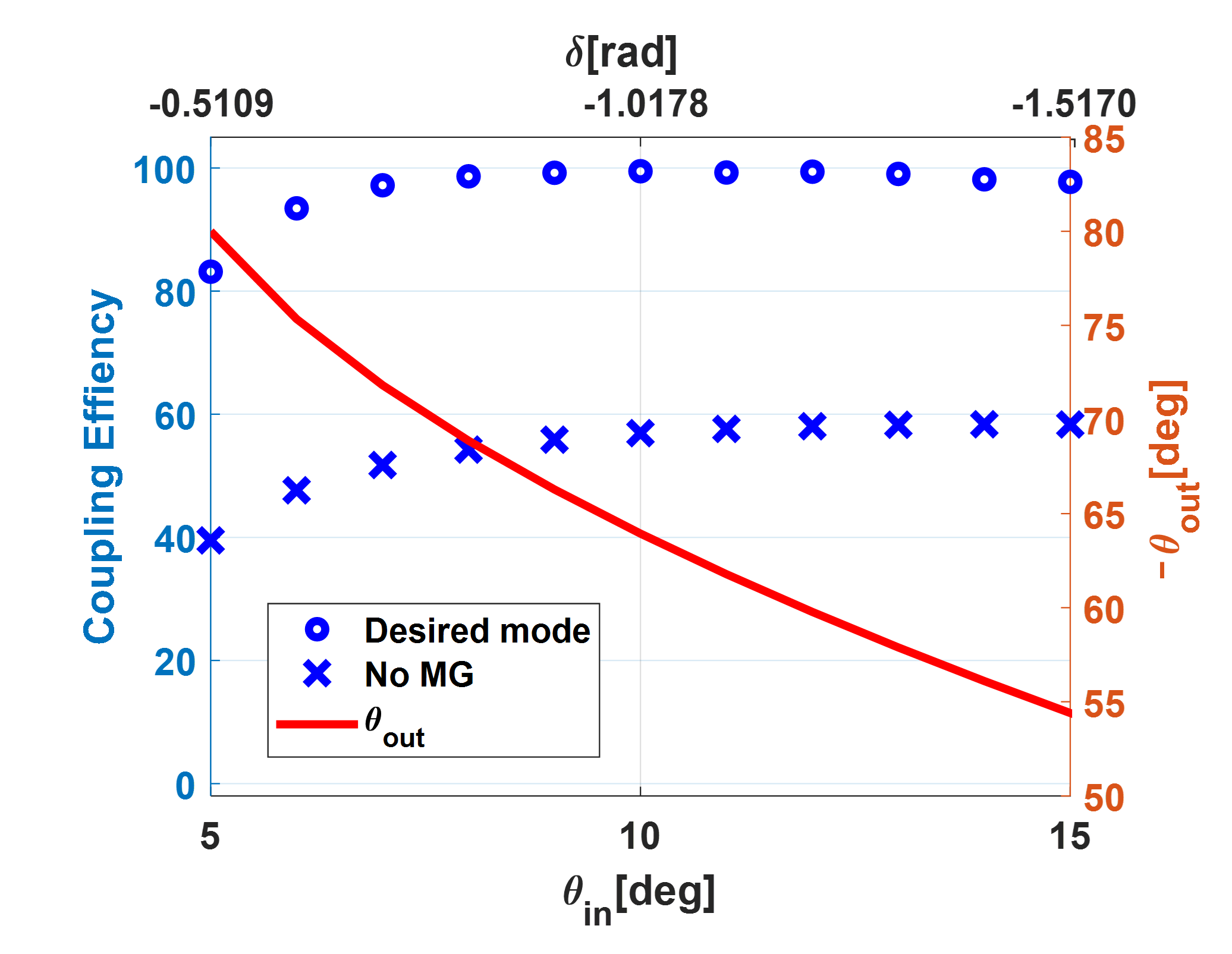}}

\caption{Coupling efficiency (blue circles, left $y$ axis) to the scanning
angle $\theta_{\text{out}}$ (red solid line, right $y$ axis) for
the designed fixed-beam phased array of Section \ref{subsec:Fixed-beam-phased-arrays} as a function of the phasing between the active array elements $\delta$ (top) and the corresponding input angle $\theta_{\text{in}}$ (bottom) for the extended scan range $5^{\circ}-20^{\circ}$.
For reference, the coupling efficiency of the bare sparse array (without the MG superstrate) towards the same angle $\theta_{\mathrm{out}}$ is also presented (blue crosses).
\label{fig:error_field}}
\end{figure}

As discussed in Section \ref{sec:Introduction}, one of the most useful features of phased arrays is their ability to scan the main beam via electronic control circuitry. However, the methodology laid out in Section \ref{sec:Theory} was used to design a sparse array without
spurious grating lobes only for a single specific interelement phasing $\delta$ (fixed-beam scenario). Once the phased array is reconfigured to a different phasing (corresponding to a different $\theta_{\text{in}}$),
the system is not guaranteed to function properly (namely, grating lobes may emerge and deteriorate the directivity). Indeed, as can be observed in Fig. \ref{fig:error_field} for the MG designed and simulated in Section \ref{subsec:Fixed-beam-phased-arrays},
when the phasing is modified within the range $\delta\in\left[-0.51,-1.51\right]\,\text{rad}$
($\theta_{\text{in}}\in\left[5^{\circ},15^{\circ}\right]$, $\theta_{\text{out}}\in\left[-80^{\circ},-54.4^{\circ}\right]$), spurious lobes emerge.
In particular, when attempting scanning to large oblique angles $\theta_\mathrm{out}\rightarrow 80^\circ$ ($\delta\rightarrow-0.51\,\text{rad}$), as much as $17\%$ of the excitation power is radiated to undesired directions, deteriorating the overall array performance in this important operation regime.

Hence, a modification to the design scheme should be made, adapting the proposed solution as to fit also sparse arrays with scanning capabilities. More precisely, our goal is to adjust the methodology presented in Section
\ref{sec:Theory} as to devise a single static and passive MG structure
to suppress grating lobes for a range of phasing values that may be
dynamically tuned at the active array level. The key to achieve this
extended feature is to realize that the synthesis procedure described
and demonstrated in Section \ref{subsec:Fixed-beam-phased-arrays}
utilizes only two degrees of freedom for the MG superstrate design.
In particular, we chose arbitrarily from a range of possible values
for the array offsets $h_{s}$ and $d_{s}$, and the only parameters
that were dictated by the formalism were the MG-PEC distance $h$
and the load impedance $\tilde{Z}$ (effectively, the capacitor width
$W$). Thus, instead of arbitrarily setting $h_{s}$ and $d_{s}$,
we can harness them to enforce additional constraints on the system
(\textcolor{black}{achieving} multifunctionality). 

To this end, we first define two desired radiation
angles $\theta_{\text{out,1}}$ and $\theta_{\text{out,2}}$ to which
we wish to scan to using the reconfigurable active array, and deduce,
from the (sparse) interelement spacing $\Lambda$, the corresponding
phasing values $\delta_{1}$ and $\delta_{2}$. Subsequently, we
consider a range of array offset combinations ($h_{s}$, $d_{s}$),
and for each of these combinations, we calculate following \eqref{eq:condition}
and \eqref{eq:impedance} the required MG vertical coordinate $h_{1}$
and the load impedance $\tilde{Z}_{1}$ that would yield perfect suppression
of the spurious lobe for the first scan angle phasing $\delta_{1}$,
and the required ($h_{2}$, $\tilde{Z}_{2}$) that would lead to optimal
directivity for the second scan angle $\delta_{2}$. Finally, since we wish to devise a \emph{single} MG configuration that would fit \textcolor{black}{\emph{both}} scan angles, we average the obtained parameters $h\triangleq\left(h_{1}+h_{2}\right)/2$, $\tilde{Z}\triangleq (\tilde{Z}_{1}+\tilde{Z}_{2})/2$ and use the analytical model to evaluate the coupling to the specified angles $\theta_{\text{out,1}}$ and $\theta_{\text{out,2}}$ with these averaged values; the best\footnote{\textcolor{black}{To maximize the model accuracy, coupling efficiency calculations in this case take into account also expected loss in the MG $|I|^{2}\delta\tilde{R}/2$ \cite{rabinovich2020arbitrary}, stemming from the resistance per unit length of the copper traces $\delta\tilde{R}=12.3\times10^{-3}\left[\eta/\lambda\right]$ estimated in \cite{epstein2017unveiling}.}} set of ($h_{s}$, $d_{s}$, $h$, $\tilde{Z}$)
in terms of coupling the desired angles found in this way is then chosen for implementation of the dynamic system.
For such ($h_{s}$, $d_{s}$), the same static passive MG configuration characterized by $h$ and $\tilde{Z}$ would be able to support grating lobe suppression for
two different scan angles. In fact, we expect that with such a MG superstrate, if the angles $\theta_{\text{in},1}$ and $\theta_{\text{in},2}$ are not too far apart, successful single-beam radiation of the combined
sparse system (array and MG) at the two phasing values would enable efficient scanning for a range of angles between them. 

In order to demonstrate the method, we choose \textcolor{black}{again an} interelement spacing
of $\Lambda=0.93\lambda$, and a desired radiation range between $\theta_{\text{out},1}=-80^{\circ}$
and $\theta_{\text{out},2}=-63.93^{\circ}$, from which phase delay
of $\delta_{1}=-0.51\text{\,rad}$ and $\delta_{2}=-1.02\text{\,rad}$
(corresponding to $\theta_{\mathrm{in},1}=5^{\circ}$ and $\theta_{\mathrm{in},2}=10^{\circ}$
respectively) is deduced from \eqref{eq:Lambda}. In accordance \textcolor{black}{with} the algorithm described above, we swept $100\times 100$ values of offset combinations ($h_{s}$, $d_{s}$) in the ranges $0<h_{s}\leq\lambda$ and $0<d_{s}\leq0.99\Lambda$, calculated for each the optimal $h$ and $\tilde{Z}$ in terms of simultaneous coupling to both scan angles, and chose the best combination. \textcolor{black}{The optimal coordinate set was found thus to be} $(h_{s},d_{s})=(0.0270\lambda,0.2125\lambda)$, implying {[}through \eqref{eq:condition} and \eqref{fig:Normalized Ploss}{]} that $h=0.2324\lambda$ and $\tilde{Z}=-6.32j\left[\eta/\lambda\right]$ \textcolor{black}{[}effectively, $W=84.81\,\text{mil}$ via \eqref{eq:capWidth}\textcolor{black}{]}. 

\begin{figure}
\centerline{\includegraphics[width = 9 cm]{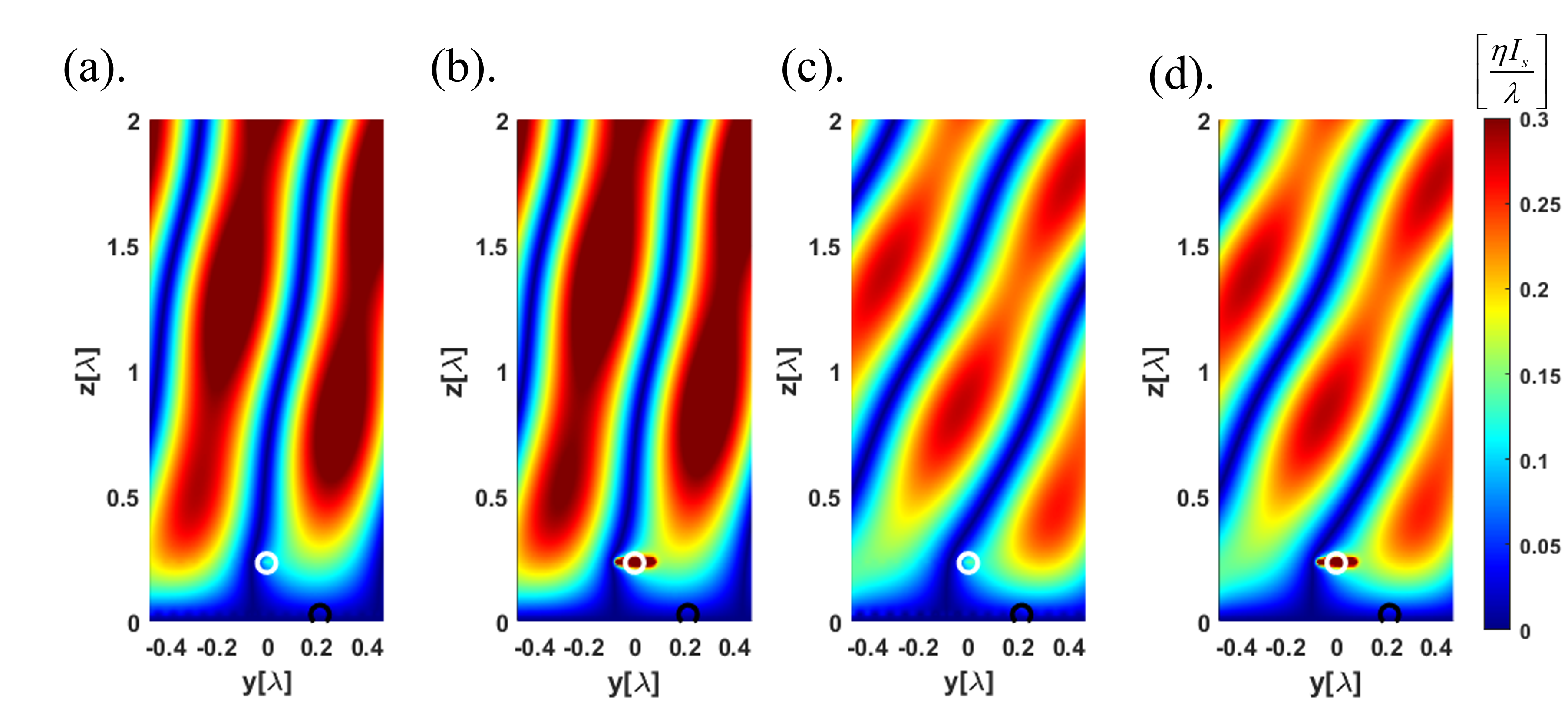}} 

\caption{Electric-field distribution $\left|\Re\left\{ E_{x}\left(y,x\right)\right\} \right|$
over a single $\Lambda$-sized period as obtained from the (a),(c)
analytical model and (b),(d) the full-wave simulation of the designed
scanning-capable sparse antenna array, tested for (a)-(b) $\theta_{\text{in},1}=5^{\circ},\,\theta_{\text{out},1}=-80^{\circ}$ and for (c)-(d) $\theta_{\text{in},2}=10^{\circ},\,\theta_{\text{out},2}=-63.93^{\circ}$.
The active electric line sources and the MG passive conducting wire are marked in black and white respectively. \label{fig:Two-angles-field}}
\end{figure}

To validate the proposed method for designing sparse scanning antenna
arrays, we plot in Fig. \ref{fig:Two-angles-field} field snapshots
corresponding to the two considered active element phasing scenarios:
one leading to radiation towards $\theta_{\text{out},1}=-80^{\circ}$
[Fig. \ref{fig:Two-angles-field}(a)-(b)], and the other radiating towards $\theta_{\text{out},2}=-63.93^{\circ}$ [Fig. \ref{fig:Two-angles-field}(c)-(d)].
As can be seen, even though we use the same static passive MG configuration (including trace geometry),
the sparse array can achieve grating-lobe-free radiation to both angles simultaneously.
Excellent agreement is again observed between the results from the analytical model and the ones obtained from full-wave simulation. 

\begin{figure}
\centerline{\includegraphics[width = 7 cm]{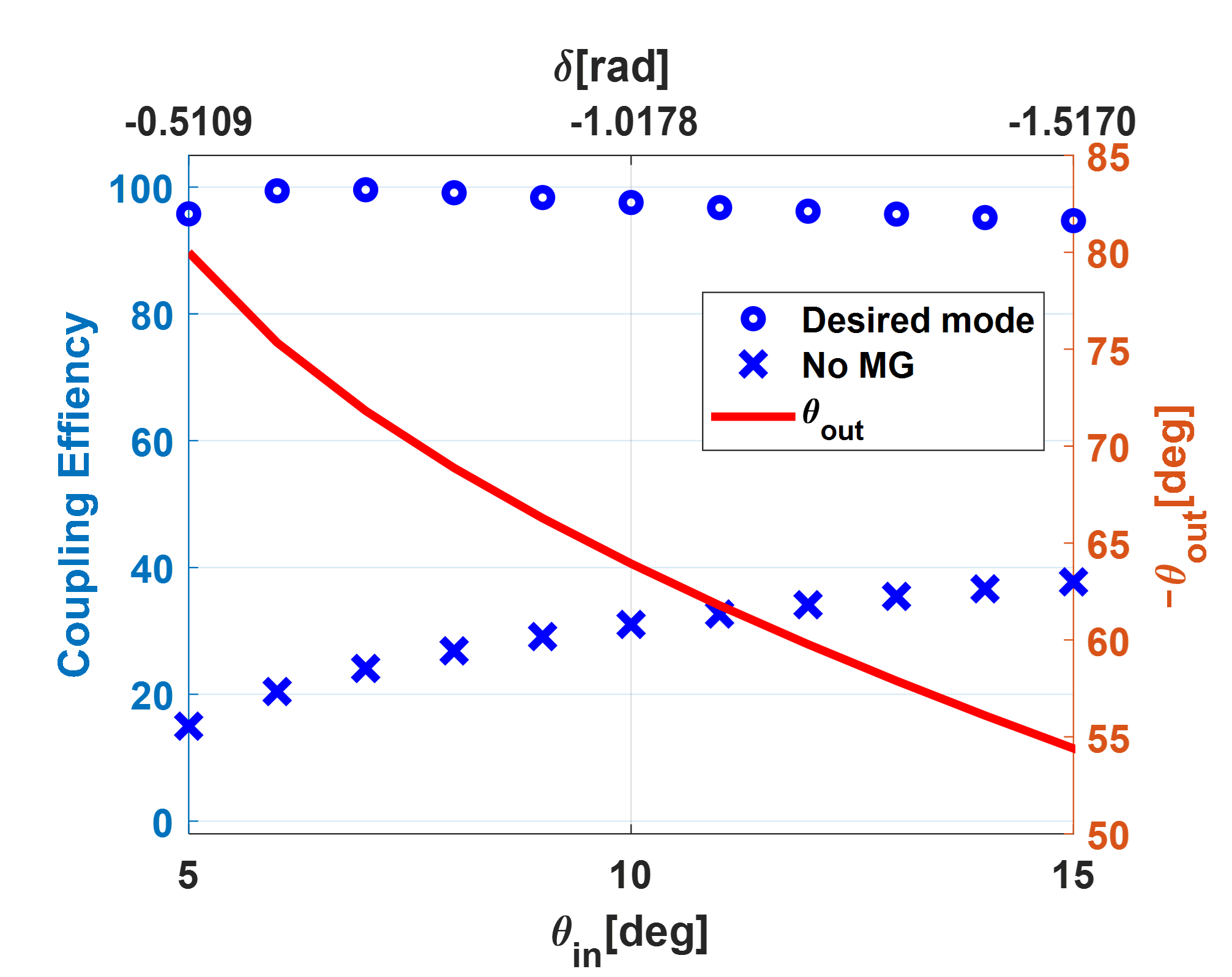}} 

\caption{Coupling efficiency (blue circles, left $y$ axis) to the scanning
angle $\theta_{\text{out}}$ (red solid line, right $y$ axis) for
the designed scanning structure as functions of the phasing between
the active array elements $\delta$ (top) and the corresponding input
angle $\theta_{\text{in}}$ (bottom) for the extended scan range $5^{\circ}-20^{\circ}$.
The coupling efficiency of the bare sparse array (without the MG superstrate)
towards the same angle $\theta_{out}$ is also presented (blue crosses).
\label{fig:Coupling-efficiency}}
\end{figure}

In fact, when considering phasing corresponding to scanning across
the whole range of angles between $\theta_{\text{out},1}$ and $\theta_{\text{out},2}$,
simulations reveal that the sparse MG-enhanced array continues to
function very well. As can be seen in Fig. \ref{fig:Coupling-efficiency},
the scanning even extends further from this range, allowing highly-directive
radiation (above 98\% coupling to the desired mode, in the presence of realistic losses) from $\theta_{\text{out}}=-60^{\circ}$ to $-80^{\circ}$
(blue circles). This result is especially impressive if one recalls that the design scheme included no full-wave optimization whatsoever, eventually leading to a detailed MG design that can suppress grating lobes for a wide range of scan angles in an array with interelement spacing of $\Lambda=0.93\lambda$, significantly larger than the conventional limit of $\lambda/2$. Indeed, without the \textquotedblright corrective\textquotedblright{}
MG superstrate, the sparse nature of the array reduces substantially the directivity, with coupling efficiencies to $\theta_{out}$ dropping by 60\%-80\% due to the untreated grating lobes
(blue crosses in Fig. \ref{fig:Coupling-efficiency}).

Another compelling benefit of the design is that it allows significant
expansion of the range of scanning angles: while the phasing in the
active array is quite mild, corresponding originally to a moderate
scan range of $\textcolor{black}{7}^{\circ}$ in $\theta_{\text{in}}$ (\textcolor{black}{from} $\theta_{\mathrm{in}}=5{}^{\circ}$
to $\theta_{\mathrm{in}}=\textcolor{black}{12}^{\circ}$), the combined system (active array
and passive MG superstrate) actually radiates across a $\textcolor{black}{20}^{\circ}$
angular range (for $\theta_{\mathrm{out}}=-80{}^{\circ}$ to $\theta_{\mathrm{out}}=-\textcolor{black}{60}{}^{\circ}$),
more than 2.5 times larger.

These results confirm our hypothesis from the beginning of this subsection, indicating that the proposed system indeed features sufficient degrees of freedom to sustain near-optimal directivity across a substantial angular range.

\subsection{Finite arrays}

\label{subsec:Finite-Arrays}

\begin{figure}
\centerline{\includegraphics[width = 9 cm]{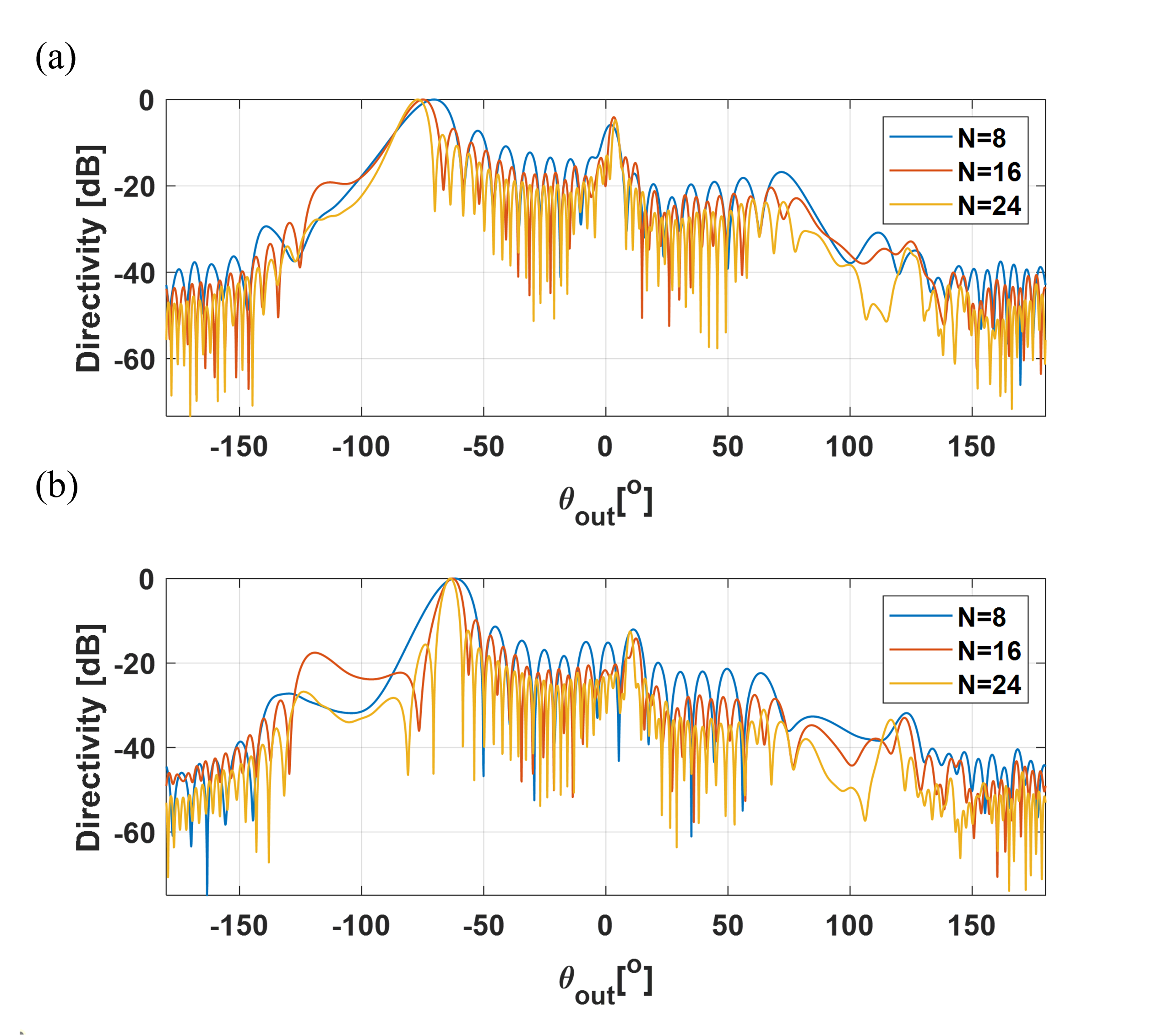}} 

\caption{Full-wave simulated normalized radiation patterns (in decibels) of the designed scanning sparse antenna array presented in Fig. \ref{fig:Two-angles-field}-\ref{fig:Coupling-efficiency} when truncated to consist of N=8,16, and 24 elements, when scanning towards (a) $\theta_{out}=-80^{\circ}$
(b) $\theta_{out}=-63.93^{\circ}$. Quantitative performance figures are listed in Table \ref{tab:effective-length}.\label{fig:radiation-patt}}
\end{figure}

While the analytical model presented in Section \ref{sec:Theory}
and the derived design schemes demonstrated in Section \ref{subsec:Fixed-beam-phased-arrays} and \ref{subsec:Dynamic-scanning} have considered an ideal periodic
array of infinite extent, in practical scenarios, finite arrays will
be used. Nonetheless, since the synthesis method was derived for infinite structures, it would be reasonable to expect it should work well for sufficiently-large arrays\textcolor{black}{, where edge effects are minor}. Correspondingly, we consider in Fig. \ref{fig:radiation-patt} the performance of the combined scanning system (sparse antenna array and MG superstrate) designed and analyzed in Section \ref{subsec:Dynamic-scanning}, as a function of the number of elements in the array. In particular, we consider three different configurations, with $N=8$, $N=16$, and
$N=24$ elements for both the active array and the MG, and evaluate using full-wave simulations\footnote{\textcolor{black}{The finite arrays were simulated using} radiation boundary conditions, with a bounding box of size  $L \times (N\Lambda+4\lambda) \times 3\lambda$ surrounding the defined MG-enhanced array configuration.}
the expected radiation patterns and directivity of these finite versions of the system, for two scan angles within the valid range (Fig. \ref{fig:Coupling-efficiency}).

The radiation patterns (Fig. \ref{fig:radiation-patt}) indeed indicate that the grating lobe suppression capability demonstrated by the infinite periodic configuration is retained also for the finite system scenario (finite array and finite MG superstrate), with the main beam clearly pointing out towards the required near-grazing angle. As one may expect, the performance is improved as the number of elements (array length)
grows, leading to a narrower beam (larger aperture) and better spurious lobe suppression (closer to the model used for the MG design). It is important to note that the difference in gain between the main and grating lobe may not \textit{a priori} seem impressive, but if one considers the power delivered within each one of the beams (i.e., including the beamwidths as well, and not only the peak directivity), it becomes apparent that a negligible fraction of the radiated power propagates to undesired angular regions.

These observations are clearly reflected in the quantified figures of merit of the MG-enhanced array presented in Table \ref{tab:effective-length}. Indeed, it is seen that as the array size increases, the aperture illumination efficiency, measuring the device directivity with respect to an ideal uniform array of the same size radiating with a single main beam towards the same $\theta_\mathrm{out}$ angle, approaches the optimal 100$\%$. This serves as an ultimate evidence to the successful suppression of the grating lobes in the devised sparse array configurations.

Another feature that is worth mentioning relates to the actual observation angle where the maximal directivity is obtained for the finite arrays. Since large oblique angles of radiation suffer more from the reduced
effective aperture, with the directivity decreasing as $\cos\theta_{\mathrm{out}}$
\cite{balanis2016antenna}, the simulated (actual) main beam angle peak $\theta_{\mathrm{out}}$
undergoes a shift towards broadside \cite{chen2018theory}. However,
as can be seen from the table, as the total array length increases,
the beamwidth decreases \cite{balanis2016antenna}, which reduces
the effects of this $\cos\theta_{\mathrm{out}}$ taper on the beam shape, yielding a
peak angle closer to the designated $\theta_{\text{out}}$. Overall, these results verify that our analytical-model based sparse array synthesis method is indeed effective for practical applications, reaching the predicted performance improvement for large-enough finite arrays as well.

\begin{table*}
\caption{\textcolor{black}{Performance figures for finite versions of the MG-enhanced antenna arrays designed in Section \ref{subsec:Dynamic-scanning}.}
\label{tab:effective-length}}
\centerline{%
\begin{tabular}{|c|c|c||c|c|c|}
\hline 
Number of elements & Total length & Desired $\theta_{\mathrm{out}}$ & Actual $\theta_{\mathrm{out}}$ & 2D Directivity [dBi] \cite{lovat2006analysis} & Aperture illumination efficiency\tabularnewline
\hline 
\hline 
8 & $6.53\lambda$ & $-80^{\circ}$ & $-71.5^{\circ}$ & $12.21$ & $78\%$\tabularnewline
\hline 
16 & $13.99\lambda$ & $-80^{\circ}$ & $-74.6^{\circ}$ & $13.55$ & $102\%$\tabularnewline
\hline 
24 & $21.46\lambda$ & $-80^{\circ}$ & $-76.7^{\circ}$ & $14.77$ & $100\%$\tabularnewline
\hline 
\hline
8 & $6.53\lambda$ & $-63.93^{\circ}$ & $-63.7^{\circ}$ & $13.66$ & $78\%$\tabularnewline
\hline 
16 & $13.99\lambda$ & $-63.93^{\circ}$ & $-62.7^{\circ}$ & $15.95$ & $102\%$\tabularnewline
\hline 
24 & $21.46\lambda$ & $-63.93^{\circ}$ & $-63.5$ & $17.79$ & $100\%$\tabularnewline
\hline 
\end{tabular}} 
\end{table*}

\section{Conclusion}
To conclude, we have presented an analytical model \textcolor{black}{integrating active} line sources as excitations to a MG, and utilized it to \textcolor{black}{devise} a simple semianalytical design scheme for sparse, \textcolor{black}{grating-lobe-free,} uniform \textcolor{black}{MG-}enhanced antenna arrays. 
\textcolor{black}{Based on the periodic structure underlying such configurations, and the proven ability of MG composites to control the coupling to scattered FB modes, single-beam radiation can be obtained by tuning the location and geometry of the meta-atoms (loaded wires) as to generate destructive interference in undesired directions.} 

The model was developed \textcolor{black}{first} for fixed beam array\textcolor{black}{s, and extended as a second step} to enable also scanning capabilities\textcolor{black}{, utilizing the available degrees of freedom to synthesize a single passive MG superstrate that could effectively suppress spurious lobes within a range of scan angles}. \textcolor{black}{As confirmed via full-wave simulations, this solution works well also when} 
a more realistic case with a finite number of elements was \textcolor{black}{considered, reaching the ideal unitary aperture illumination efficiency for sufficiently large (but finite) arrays.} 
\textcolor{black}{The presented methodology, yielding a semianalytically designed PCB-compatible structure that can be attached modularly to conventional antenna arrays, may provide an elegant and effective path to reduce design complexity and cost of dynamic beam steering antennas required in next generation communication systems.}

\end{document}